\documentclass[12pt]{article}
\usepackage{graphicx}
\def\beq{\begin{equation}}
\def\eeq{\end{equation}}
\def\bea{\begin{eqnarray}}
\def\eea{\end{eqnarray}}
\def\nn{\nonumber}

\def\roughly#1{\mathrel{\raise.3ex\hbox
{$#1$\kern-.75em\lower1ex\hbox{$\sim$}}}}

\def\sla#1{\raise.15ex\hbox{$/$}\kern-.57em #1}

\def\ks{K_S}
\def\kbar{{\bar K}^0}
\def\bd{B_d^0}
\def\bdbar{{\bar B}_d^0}
\def\pewc{P_{EW}^C}
\def\pew{P_{EW}}

\def\btos{{\bar b} \to {\bar s}}
\def\order{\lower 1.8ex \hbox{\LARGE\~{}}}
\def\btokpi{B \to K \pi}
\def\btokpipi{B \to K \pi \pi}
\def\btokkk{B \to KK{\bar K}}

\makeatletter
\newcommand{\contraction}[5][1ex]{%
 \mathchoice
  {\contraction@\displaystyle{#2}{#3}{#4}{#5}{#1}}%
  {\contraction@\textstyle{#2}{#3}{#4}{#5}{#1}}%
  {\contraction@\scriptstyle{#2}{#3}{#4}{#5}{#1}}%
  {\contraction@\scriptscriptstyle{#2}{#3}{#4}{#5}{#1}}}%
\newcommand{\contraction@}[6]{%
 \setbox0=\hbox{$#1#2$}%
 \setbox2=\hbox{$#1#3$}%
 \setbox4=\hbox{$#1#4$}%
 \setbox6=\hbox{$#1#5$}%
 \dimen0=\wd2%
 \advance\dimen0 by \wd6%
 \divide\dimen0 by 2%
 \advance\dimen0 by \wd4%
 \vbox{%
  \hbox to 0pt{%
   \kern \wd0%
   \kern 0.4\wd2
   \contraction@@{\dimen0}{#6}%
   \hss}%
  \vskip 0.8ex
  \vskip\ht2}}

\newcommand{\contraction@@}[3][0.06em]{%
 \hbox{%
  \vrule width #1 height 0pt depth #3%
  \vrule width #2 height 0pt depth #1%
  \vrule width #1 height 0pt depth #3%
  \relax}}
\makeatother
\begin{document}

\begin{flushright}
UdeM-GPP-TH-11-203 \\
\end{flushright}

\begin{center}
\bigskip
{\Large \bf \boldmath Measuring $\gamma$ with $\btokpipi$ and $\btokkk$ Decays} \\
\bigskip
\bigskip
{\large 
Nicolas Rey-Le Lorier $^{a,}$\footnote{nr323@cornell.edu}
and David London $^{b,}$\footnote{london@lps.umontreal.ca}
}
\end{center}

\begin{flushleft}
~~~~~~~~~~~$a$: {\it Cornell University, Ithaca, NY 14853, USA}\\
~~~~~~~~~~~$b$: {\it Physique des Particules, Universit\'e
de Montr\'eal,}\\
~~~~~~~~~~~~~~~{\it C.P. 6128, succ. centre-ville, Montr\'eal, QC,
Canada H3C 3J7}\\
\end{flushleft}

\begin{center}
\bigskip (\today)
\vskip0.5cm {\Large Abstract\\} \vskip3truemm
\parbox[t]{\textwidth}{We present a method for cleanly extracting the
  CP phase $\gamma$ from the Dalitz plots of $B^+ \to K^+\pi^+\pi^-$,
  $\bd \to K^+\pi^0\pi^-$, $\bd \to K^0\pi^+\pi^-$, $\bd \to K^+ K^0
  K^-$, and $\bd \to K^0 K^0 \kbar$. The $\btokpipi$ and $\btokkk$
  decays are related by flavor SU(3) symmetry, but SU(3) breaking is
  taken into account. Most of the experimental measurements have
  already been made -- what remains is a Dalitz-plot analysis of $\bd
  \to K^0 K^0 \kbar$ (or $\bd\to\ks\ks\ks$ ). We (very) roughly
  estimate the error on $\gamma$ to be $\sim 25\%$. This is somewhat
  larger than the error in two-body decays, but it would be the first
  clean measurement of $\gamma$ in three-body decays. Furthermore, at
  the super-$B$ factory, it is possible that $\gamma$ could be
  measured more precisely in three-body decays than in two-body
  decays.}
\end{center}


\thispagestyle{empty}
\newpage
\setcounter{page}{1}
\baselineskip=14pt

In the past, most of the theoretical work looking at clean methods for
extracting weak-phase information in the $B$ system focused on
two-body decays. This is essentially because (i) final states such as
$\psi \ks$, $\pi^+\pi^-$, etc.\ are CP eigenstates, and (ii) if there
is a second decay amplitude, with a different weak phase, it has been
possible to find methods to remove this ``pollution,'' and cleanly get
at the weak phases. On the other hand, in three-body $B$ decays, final
states such as $\ks \pi^+\pi^-$ are not CP eigenstates -- the value of
its CP depends on whether the relative $\pi^+\pi^-$ angular momentum
is even (CP $+$) or odd (CP $-$). Furthermore, even if the CP of the
final state were determined in some way, one still has the problem of
removing the pollution due to additional decay amplitudes. For these
reasons, it has generally been thought that it is not possible to
obtain clean weak-phase information from three-body decays
\cite{CKMconstraints}.

Recently, it was shown that this is not true.  By doing a diagrammatic
analysis of the three-body amplitudes, one can resolve these two
problems \cite{diagramspaper}.  First, a Dalitz-plot analysis can be
used to experimentally separate the CP $+$ and $-$ components of the
three-particle final state.  Second, one can often remove the
pollution of additional diagrams and cleanly measure the CP phases. In
fact, in Ref.~\cite{Kpipipaper}, it was shown how to extract the weak
phase $\gamma$ from $\btokpipi$ decays. We briefly describe this
method below.

In $\btokpipi$ decays, the isospin state of the $\pi\pi$ pair must be
symmetric (antisymmetric) if the relative angular momentum is even
(odd). As we will see below, it is the symmetric case which is most
interesting. Here there are six possible decays: $B^+ \to
K^+\pi^+\pi^-$, $B^+ \to K^+\pi^0\pi^0$, $B^+ \to K^0\pi^+\pi^0$, $\bd
\to K^+\pi^-\pi^0$, $\bd \to K^0\pi^+\pi^-$, and $\bd \to
K^0\pi^0\pi^0$. The first step is to express the amplitudes for these
processes in terms of diagrams. The diagrams are as in two-body $B$
decays \cite{GHLR}: the color-favored and color-suppressed tree
amplitudes $T$ and $C$, the gluonic-penguin amplitudes $P_{tc}$ and
$P_{uc}$, and the color-favored and color-suppressed
electroweak-penguin (EWP) amplitudes $\pew$ and $\pewc$. (We neglect
annihilation- and exchange-type diagrams.) Furthermore, for three-body
decays, it is necessary to ``pop'' a quark pair from the vacuum.  The
diagrams are written with subscripts, indicating that the popped quark
pair is between two (non-spectator) final-state quarks (subscript
`1'), or between two final-state quarks including the spectator
(subscript `2'). (For $\btokpipi$ decays, the popped quark pair is
$u{\bar u}$ or $d{\bar d}$. Under isospin, these amplitudes are
equal.)

In addition, some time ago it was shown that, under flavor SU(3)
symmetry, there are relations between the EWP and tree diagrams in
$\btokpi$ decays \cite{NR,GPY}. In Ref.~\cite{Kpipipaper}, it was
shown that similar EWP-tree relations hold for $\btokpipi$
decays. Taking $c_1/c_2 = c_9/c_{10}$ for the Wilson coefficients
(which holds to about 5\%), these take the simple form
\bea
P'_{EW1} = \kappa T'_1~,~~~~~~
P'_{EW2} =\kappa T'_2~,\nn\\
P^{\prime C}_{EW1} = \kappa C'_1~,~~~~~~
P^{\prime C}_{EW2} = \kappa C'_2~,
\label{EWPtree}
\eea
where
\bea
\kappa \equiv - \frac{3}{2} \frac{|\lambda_t^{(s)}|}{|\lambda_u^{(s)}|} \frac{c_9+c_{10}}{c_1+c_2} ~,
\eea
with $\lambda_p^{(s)}=V^*_{pb} V_{ps}$.

Now, the EWP-tree relations assume SU(3) symmetry (and the approximate
ratio of Wilson coefficients).  The expected error due to
SU(3)-breaking effects is $O(30\%)$. However, the dominant diagram in
$\btos$ decays is $P'_{tc}$, so that EWPs and trees are subleading
effects. Thus SU(3) breaking is subdominant -- the net theoretical
error due to the use of the EWP-tree relations is only $O(5\%)$.  This
is consistent with the error estimates given in Ref.~\cite{NR} (for
EWP-tree relations in $\btokpi$).

In addition, there is an important caveat. Under SU(3), the final
state in $\btokpipi$ involves three identical particles, so that the
six permutations of these particles (the group $S_3$) must be taken
into account.  That is, the three particles are in a totally symmetric
state, a totally antisymmetric state, or one of four mixed states.
However, the EWP-tree relations hold only for the totally symmetric
state. Thus, the analysis must be carried out for this state. Now, the
expressions for the $B \to K(\pi\pi)_{sym}$ amplitudes in terms of
diagrams hold even under full SU(3) symmetry \cite{Kpipipaper}. It is
therefore only necessary to produce observables for the totally
symmetric states. This is doable, and below we present the details of
how this is carried out.

With the above EWP-tree relations, the six $B \to K(\pi\pi)_{sym}$
amplitudes can be written in terms of 5 effective diagrams
(i.e.\ linear combinations of the diagrams) \cite{Kpipipaper}.  There
are therefore 10 theoretical parameters in the amplitudes\footnote{In
  fact, the expression for any indirect CP asymmetry contains another
  theoretical parameter -- the phase of $\bd$-$\bdbar$ mixing,
  $\beta$. However, its value can be taken from the indirect CP
  asymmetry in $\bd\to J/\psi \ks$ \cite{pdg}.}: 5 magnitudes of
effective diagrams, 4 relative (strong) phases, and $\gamma$. On the
other hand, there are 11 experimental observables. Given that $B^+ \to
K^0\pi^+\pi^0$ is not independent (its amplitude is proportional to
that of $\bd \to K^+\pi^0\pi^-$), these are the branching ratios and
direct CP asymmetries of $B^+ \to K^+\pi^+\pi^-$, $B^+ \to
K^+\pi^0\pi^0$, $\bd \to K^+\pi^0\pi^-$, $\bd \to K^0\pi^+\pi^-$, and
$\bd \to K^0\pi^0\pi^0$, and the indirect CP asymmetry of $\bd \to
K^0\pi^+\pi^-$ (the indirect CP asymmetry of $\bd \to K^0\pi^0\pi^0$
will essentially be impossible to measure).  Since there are more
observables than theoretical parameters, $\gamma$ can be extracted by
doing a fit\footnote{There is a complication in that the diagrams are
  momentum dependent, as are the observables.  In obtaining the
  best-fit ``values'' of the diagrams, one will determine the momentum
  dependence of their magnitudes and relative strong phases. On the
  other hand, $\gamma$ is independent of the particles' momenta. Later
  in the paper, we detail how such a fit is done.}.

The disadvantage of this method is that it involves the decays $B^+
\to K^+\pi^0\pi^0$ and $\bd \to K^0\pi^0\pi^0$. With two $\pi^0$
mesons in the final state, both of these decays will be extremely
difficult to measure. We are therefore motivated to see if the
$\btokpipi$ method can be modified, avoiding these two decays. As we
show below, this can indeed be done -- things can be considerably
improved by using $\btokkk$ decays. The use of these decays is quite
natural since they, like $\btokpipi$, are also $\btos$ transitions.

First, consider $\btokpipi$ decays with the $\pi\pi$ pair in a
symmetric isospin state. We leave aside $B^+ \to K^+\pi^0\pi^0$, $\bd
\to K^0\pi^0\pi^0$ and $B^+ \to K^0\pi^+\pi^0$ (since, as mentioned
above, its amplitude is not independent). The amplitudes of the
remaining three processes are
\bea
2 A(\bd \to K^+\pi^0\pi^-)_{sym} &=& T'_1 e^{i\gamma}+C'_2 e^{i\gamma} - P'_{EW2} - P^{\prime C}_{EW1} ~, \nn\\
\sqrt{2} A(\bd \to K^0\pi^+\pi^-)_{sym} &=& -T'_1 e^{i\gamma}-C'_1 e^{i\gamma}-{\tilde P}'_{uc} e^{i\gamma}+ {\tilde P}'_{tc} \nn\\
&& \hskip1.5truecm +~\frac13 P'_{EW1} + \frac23 P^{\prime C}_{EW1} - \frac13 P^{\prime C}_{EW2} ~, \nn\\
\sqrt{2} A(B^+ \to K^+\pi^+\pi^-)_{sym} &=& -T'_2 e^{i\gamma}-C'_1 e^{i\gamma}-{\tilde P}'_{uc} e^{i\gamma}+ {\tilde P}'_{tc} \nn\\
&& \hskip1.5truecm +~\frac13 P'_{EW1} - \frac13 P^{\prime C}_{EW1} + \frac23 P^{\prime C}_{EW2} ~.
\label{kpipiamps}
\eea
In the above, ${\tilde P}' \equiv P'_1 +P'_2$.  (As $\btokpipi$ is a
$\btos$ transition, the diagrams are written with primes.) Here we
have explicitly written the weak-phase dependence (this includes
$\gamma$ and the minus sign from $V_{tb}^* V_{ts}$ [${\tilde P}'_{tc}$
  and EWPs]), while the diagrams contain strong phases.

Second, consider $\btokkk$ decays. For the case in which the final
$KK$ pair is in a symmetric isospin state, there are four such
processes: $B^+ \to K^+ K^+ K^-$, $B^+ \to K^+ K^0 \kbar$, $\bd \to
K^+ K^0 K^-$, and $\bd \to K^0 K^0 \kbar$. Here, $B^+ \to K^+ K^+ K^-$
and $B^+ \to K^+ K^0 \kbar$ are not independent -- their amplitudes
are proportional to those of $\bd \to K^+ K^0 K^-$ and $\bd \to K^0
K^0 \kbar$, respectively.  These are
\bea
\label{kkkamps}
\sqrt{2} A(\bd \to K^+ K^0 K^-)_{sym} &=& -T'_{2,s} e^{i\gamma}-C'_{1,s} e^{i\gamma}
-{\hat P}'_{uc} e^{i\gamma}+ {\hat P}'_{tc} \nn\\
&& \hskip0.8truecm +~\frac23 P'_{EW1,s} - \frac13 P'_{EW1} + \frac23
P^{\prime C}_{EW2,s} - \frac13 P^{\prime C}_{EW1} ~, \nn\\
A(\bd \to K^0 K^0 \kbar)_{sym} &=& {\hat P}'_{uc} e^{i\gamma}- {\hat P}'_{tc} \\
&& \hskip0.8truecm +~\frac13 P'_{EW1,s} + \frac13 P'_{EW1} + \frac13
P^{\prime C}_{EW2,s} + \frac13 P^{\prime C}_{EW1} ~, \nn
\eea
where ${\hat P}' \equiv P'_{2,s} + P'_1$.  In the above, certain
diagrams are written with the subscript `$s$.' This indicates that the
popped quark pair is $s{\bar s}$.  When the diagram has no subscript
$s$ (the penguin or EWP diagrams), this means that the popped quark
pair is $u{\bar u}$ or $d{\bar d}$, but the virtual particle decays to
$s{\bar s}$.

We now assume flavor SU(3) symmetry. This has two consequences. First,
the amplitude with a popped $s{\bar s}$ quark pair is equal to that
with a popped $u{\bar u}$ or $d{\bar d}$. That is, we no longer need
the subscript $s$ on diagrams. This means that the diagrams in
$\btokkk$ decays are the same as those in $\btokpipi$ decays.  Second,
the EWP-tree relations of Eq.~(\ref{EWPtree}) hold.

Thus, under SU(3) the amplitudes of Eqs.~(\ref{kpipiamps}) and
(\ref{kkkamps}) take the form
\bea
2 A(\bd \to K^+\pi^0\pi^-)_{sym} &=& T'_1 e^{i\gamma}+C'_2 e^{i\gamma} - \kappa \left(T'_2 + C'_1\right) ~, \nn\\
\sqrt{2} A(\bd \to K^0\pi^+\pi^-)_{sym} &=& -T'_1 e^{i\gamma}-C'_1 e^{i\gamma}-{\tilde P}'_{uc} e^{i\gamma}+ {\tilde P}'_{tc} \nn\\
&& \hskip1.5truecm +~\kappa \left(\frac13 T'_1 + \frac23 C'_1 - \frac13 C'_2\right) ~, \nn\\
\sqrt{2} A(B^+ \to K^+\pi^+\pi^-)_{sym} &=& -T'_2 e^{i\gamma}-C'_1 e^{i\gamma}-{\tilde P}'_{uc} e^{i\gamma}+ {\tilde P}'_{tc} \nn\\
&& \hskip1.5truecm +~\kappa \left(\frac13 T'_1 - \frac13 C'_1 + \frac23 C'_2 \right) ~, \nn\\
\sqrt{2} A(\bd \to K^+ K^0 K^-)_{sym} &=& -T'_2 e^{i\gamma}-C'_1 e^{i\gamma}
-{\tilde P}'_{uc} e^{i\gamma}+ {\tilde P}'_{tc} \nn\\
&& \hskip0.8truecm +~\kappa \left( \frac13 T'_1 - \frac13 C'_1  + \frac23
C'_2 \right) ~, \nn\\
A(\bd \to K^0 K^0 \kbar)_{sym} &=& {\tilde P}'_{uc} e^{i\gamma}- {\tilde P}'_{tc} \nn\\
&& \hskip0.8truecm +~\kappa \left(\frac23 T'_1 + \frac13 C'_1 + \frac13
C'_2 \right) ~. 
\eea
Note that this implies that $A(B^+ \to K^+\pi^+\pi^-)_{sym} = A(\bd
\to K^+ K^0 K^-)_{sym}$.  Further, we reiterate that the above
expressions for the amplitudes hold also for the totally symmetric
final state, to which the EWP-tree relations apply.

We now define the following five effective diagrams:
\bea
T'_a &\equiv& T'_1 - T'_2 ~,\nn\\
T'_b &\equiv& C'_2 + T'_2 ~,\nn\\
P'_a &\equiv& {\tilde P}'_{uc} + T'_2 + C'_1 ~,\nn\\
P'_b &\equiv& {\tilde P}'_{tc} + \kappa \left( \frac13 T'_1 + \frac23 C'_1 - \frac13 C'_2 \right) ~,  \nn\\
C'_a &\equiv& \kappa \left(C'_1 - C'_2 \right) ~. 
\label{eq:effdiag}
\eea
The amplitudes can be written in terms of these five diagrams:
\bea
\label{effamps}
2 A(\bd \to K^+\pi^0\pi^-)_{sym} &=& T'_a e^{i\gamma} + T'_b e^{i\gamma} - C'_a - \kappa T'_b ~, \nn\\
\sqrt{2} A(\bd \to K^0\pi^+\pi^-)_{sym} &=& - T'_a e^{i\gamma} - P'_a e^{i\gamma} + P'_b ~, \nn\\
\sqrt{2} A(B^+ \to K^+\pi^+\pi^-)_{sym} &=& - P'_a e^{i\gamma} + P'_b - C'_a ~, \nn\\
\sqrt{2} A(\bd \to K^+ K^0 K^-)_{sym} &=&  - P'_a e^{i\gamma} + P'_b - C'_a ~, \nn\\
A(\bd \to K^0 K^0 \kbar)_{sym} &=& P'_a e^{i\gamma} - T'_b e^{i\gamma} - \frac{1}{\kappa}  C'_a e^{i\gamma} \nn\\
&& \hskip1truecm -~P'_b + \kappa T'_a + \kappa T'_b + C'_a~.
\eea

As with the $\btokpipi$ method, five effective diagrams corresponds to
10 theoretical parameters: 5 magnitudes of diagrams, 4 relative
phases, and $\gamma$. But there are 11 (momentum-dependent)
experimental observables: the decay rates and direct asymmetries for
the four decays $\bd \to K^+\pi^0\pi^-$, $\bd \to K^0\pi^+\pi^-$, $\bd
\to K^+ K^0 K^-$ and $\bd \to K^0 K^0 \kbar$ (we ignore $B^+ \to
K^+\pi^+\pi^-$ since its amplitude is not independent), and the
indirect asymmetries of $\bd \to K^0\pi^+\pi^-$, $\bd \to K^+ K^0 K^-$
and $\bd \to K^0 K^0 \kbar$.  With more observables than theoretical
parameters, $\gamma$ can be extracted from a fit. 

We now present the details of how the fit is carried out. Consider the
decay $B \to P_1 P_2 P_3$, in which the three pseudoscalar mesons
$P_i$ ($i=1$-3) have momenta $p_i$. From these, we can construct the
three Mandelstam variables:
\beq
s_{12} \equiv \left( p_1 + p_2 \right)^2 ~~,~~~~
s_{13} \equiv \left( p_1 + p_3 \right)^2 ~~,~~~~
s_{23} \equiv \left( p_2 + p_3 \right)^2 ~.
\eeq
These are not independent, but obey
\beq
s_{12} + s_{13} + s_{23} = m_B^2 + m_1^2 + m_2^2 + m_3^2 ~.
\eeq
Experimentally, the Dalitz plot of this decay is measured. Its events
are given in terms of two Mandelstam variables, say $s_{12}$ and
$s_{13}$.  Now, the great advantage of a Dalitz-plot analysis is that
it allows one to extract the full amplitude of the decay.  We write
\beq
{\cal M}(B \to P_1 P_2 P_3) = \sum_j c_j e^{i\theta_j} F_j(s_{12},s_{13}) ~,
\label{Kpipiamp}
\eeq
where the sum is over all decay modes (resonant and non-resonant).
$c_j$ and $\theta_j$ are the magnitude and phase of the $j$
contribution, respectively, measured relative to one of the
contributing channels. The distributions $F_j$, which depend on
$s_{12}$ and $s_{13}$, describe the dynamics of the individual decay
amplitudes, and take different (known) forms for the various
contributions. The key point is that a maximum likelihood fit over the
entire Dalitz plot gives the best values of the $c_j$ and
$\theta_j$. Thus, the decay amplitude can be obtained, up to an
overall normalization. This normalization is fixed by the constraint
of the measured partial rate \cite{pdg}:
\beq
\Gamma = \frac{1}{(2\pi)^3} \frac{1}{32 m_B^3} \int |{\cal M}|^2 ds_{12} ds_{13} ~.
\label{partialrate}
\eeq
With this, the decay amplitude ${\cal M}(s_{12},s_{13})$ is known.

As will be seen below, we rely heavily on ${\cal
  M}(s_{12},s_{13})$. In particular, we use it to obtain the
observables for the $B \to P_1 P_2 P_3$ decay. As such, the errors on
these observables come entirely from the uncertainty in ${\cal
  M}(s_{12},s_{13})$. While, as noted above, it is possible to obtain
the best-fit values of the Dalitz-plot variables $c_j$ and $\theta_j$,
there are errors associated with these values.  This is due to two
sources. First, one has the statistical error in the experimental
Dalitz plot. Second, there is a systematic uncertainty related to the
choice of the $F_j$ in Eq.~(\ref{Kpipiamp}). In addition, there is a
statistical error in the overall normalization [coming from
  Eq.~(\ref{partialrate})]. All of these must be carefully taken into
account in order to obtain conservative errors on the Dalitz-plot
variables.

As noted earlier, the EWP-tree relations hold only for the totally
symmetric SU(3) decay amplitude. But this can be found from the above:
\bea
{\cal M}_{fully~sym} & = & 
\frac{1}{\sqrt{6}} \left[ {\cal M}(s_{12},s_{13}) + {\cal M}(s_{13},s_{12}) +
  {\cal M}(s_{12},s_{23}) \right. \nn\\
&& \hskip1.5truecm \left. +~{\cal M}(s_{23},s_{12}) + {\cal M}(s_{23},s_{13}) + {\cal
    M}(s_{13},s_{23}) \right] ~.
\eea
Using this, it is possible to compute the $B \to P_1 P_2 P_3$
observables.  However, recall that the method involves a fit using the
observables from several different decays ($\bd \to K^+\pi^0\pi^-$,
$\bd \to K^0\pi^+\pi^-$, $\bd \to K^+ K^0 K^-$ and $\bd \to K^0 K^0
\kbar$). All observables must involve the same Mandelstam
variables. On the other hand, the numbering of final-state particles
is arbitrary, so that $s_{12}$ for one decay might equal $s_{13}$ for
a different decay. All of this makes it somewhat confusing to ensure
that observables in different decays have the same Mandelstam
variables. For this reason, it is useful at this stage to change
notation (but the physics is unchanged). In any decay there are three
Mandelstam variables. We define $s_{++}$, $s_+$ and $s_-$ to be the
largest, second-largest, and smallest of these, respectively. The
identities of the particles which are associated with $s_{++}$, $s_+$
and $s_-$ are irrelevant (e.g.\ $s_{++}$ can correspond to $s_{12}$,
$s_{13}$ or $s_{23}$). This is consistent with the assumption of SU(3)
and the fully symmetric decay amplitude.  With these Mandelstam
variables, we have
\bea
{\cal M}_{fully~sym} & = & 
\frac{1}{\sqrt{6}} \left[ {\cal M}(s_{++},s_+) + {\cal M}(s_+,s_{++}) +
  {\cal M}(s_{++},s_-) \right. \nn\\
&& \hskip1.5truecm \left. +~{\cal M}(s_-,s_{++}) + {\cal M}(s_-,s_+) + {\cal
    M}(s_+,s_-) \right] ~.
\label{fullysym}
\eea
Since $s_{++}$, $s_+$ and $s_-$ are not independent, this gives the
fully symmetric amplitude as a function of two Mandelstam variables,
say $s_{++}$ and $s_+$.

The observables are obtained as follows. First, one forms the totally
symmetric SU(3) decay amplitudes as in Eq.~(\ref{fullysym}) for each
$B \to P_1 P_2 P_3$ decay (${\cal M}_{fully~sym}$) and its CP
conjugate (${\bar{\cal M}}_{fully~sym}$). Second, using these, for
specific values of $s_{++}$ and $s_+$, one computes the partial rates:
\bea
\Gamma_{s_{++},s_+} & = & \frac{1}{(2\pi)^3} \frac{1}{32 m_B^3} |{\cal M}_{fully~sym}(s_{++},s_+)|^2 ~, \nn\\
{\bar\Gamma}_{s_{++},s_+} & = & \frac{1}{(2\pi)^3} \frac{1}{32 m_B^3} |{\bar{\cal M}}_{fully~sym}(s_{++},s_+)|^2 ~.
\eea
These allow the computation of the CP-averaged branching ratio and
direct CP asymmetry:
\bea
\label{BRAdir}
BR_{s_{++},s_+} & = & \frac{1}{\Gamma_B} \left( \Gamma_{s_{++},s_+} + {\bar\Gamma}_{s_{++},s_+} \right) ~, \nn \\
A_{s_{++},s_+} & = & \frac{\Gamma_{s_{++},s_+} - {\bar\Gamma}_{s_{++},s_+}}{\Gamma_{s_{++},s_+} + {\bar\Gamma}_{s_{++},s_+}} ~.
\eea
Third, for those decays in which the final state is accessible to both
$\bd$ and $\bdbar$ mesons, one has an indirect (mixing-induced) CP
asymmetry. It is given by
\beq
S_{s_{++},s_+} = {\rm Im} \left[ e^{-2i\beta} \, \frac{{\bar{\cal M}}_{fully~sym}(s_{++},s_+)}{{\cal M}_{fully~sym}(s_{++},s_+)} \right] ~.
\label{Aindir}
\eeq
As discussed earlier, in all cases, the error on the observables is
found by propogating the errors on the Dalitz-plot variables. These
include both statistical and systematic effects.

Now, given that the method assumes flavor SU(3) symmetry, one would
like to know how SU(3) breaking affects the analysis, and what is its
size.  Leaving aside the EWP-tree relations, in which SU(3)-breaking
effects are subdominant, there are two areas where the breaking may be
significant. First, under SU(3), the diagrams in $\btokkk$ and
$\btokpipi$ are the same.  Since both decays are $\btos$ transitions,
the difference between them is that $\btokkk$ decays have an $s{\bar
  s}$ quark pair in the final state, hadronizing to $K{\bar K}$, while
$\btokpipi$ decays have $u{\bar u}$ or $d{\bar d}$, hadronizing to
$\pi\pi$. This is essentially the same for each diagram. (The
SU(3)-breaking effect associated with an $s{\bar s}$ pair being popped
from the vacuum may not be exactly equal to that when $s{\bar s}$ is
produced in the decay of a virtual particle, but the difference is
small.) Thus, including SU(3) breaking, the amplitudes of
Eq.~(\ref{effamps}) can be written
\bea
\label{SU3breakamps}
2 A(\bd \to K^+\pi^0\pi^-)_{sym} &=& T'_a e^{i\gamma} + T'_b e^{i\gamma} - C'_a - \kappa T'_b ~, \nn\\
\sqrt{2} A(\bd \to K^0\pi^+\pi^-)_{sym} &=& - T'_a e^{i\gamma} - P'_a e^{i\gamma} + P'_b ~, \nn\\
\sqrt{2} A(B^+ \to K^+\pi^+\pi^-)_{sym} &=& - P'_a e^{i\gamma} + P'_b - C'_a ~, \nn\\
\sqrt{2} A(\bd \to K^+ K^0 K^-)_{sym} &=&  (1+f_{SU(3)}) \left[ - P'_a e^{i\gamma} + P'_b - C'_a \right] ~, \nn\\
A(\bd \to K^0 K^0 \kbar)_{sym} &=& (1+f_{SU(3)}) \left[ P'_a e^{i\gamma} - T'_b e^{i\gamma} - \frac{1}{\kappa}  C'_a e^{i\gamma} \right. \nn\\
&& \hskip1truecm \left. -~P'_b + \kappa T'_a + \kappa T'_b + C'_a \right]~,
\eea
where $f_{SU(3)}$ is the SU(3)-breaking factor. Second, under SU(3),
$\pi$'s and $K$'s are identical particles, so that there is no
difference between the Mandelstam variables for the processes
$\btokkk$ and $\btokpipi$. There is therefore an SU(3)-breaking effect
between the fully symmetric decay amplitudes for the two types of
decay. However, it can be included in $f_{SU(3)}$.

The addition of $f_{SU(3)}$ brings the number of unknown theoretical
parameters to 11. In principle, these can all be determined from a fit
to the 11 experimental observables, albeit with discrete
ambiguities. However, we can do better. Above it was noted that, in
the limit of perfect SU(3), $A(B^+ \to K^+\pi^+\pi^-)_{sym} = A(\bd
\to K^+ K^0 K^-)_{sym}$. This means that $f_{SU(3)}$ can be determined
by a comparison of these two decays. In particular,
\bea
\frac{\tau_+}{\tau_0} \frac{\mathcal{B}(\bd \to K^+ K^0 K^-)_{sym}}{\mathcal{B}(B^+ \to K^+\pi^+\pi^-)_{sym}} = (1+f_{SU(3)})^2 ~.
\label{relBR}
\eea

In fact, this comparison can be performed now since the decays have
been measured: $B^+ \to K^+\pi^+\pi^-$ in Ref.~\cite{K+pi-pi+}, $\bd
\to K^+ K^0 K^-$ in Ref.~\cite{K+K-K0}. Now, since the EWP-tree
relations [Eq.~(\ref{EWPtree})] have been used to derive the
expressions for the amplitudes, Eq.~(\ref{relBR}) holds only for
totally symmetric states. Using the technique described above, one can
obtain $A(B^+ \to K^+\pi^+\pi^-)_{fully~sym}$ and $A(\bd \to K^+ K^0
K^-)_{fully~sym}$.  In order to get the branching ratios, we compute
the integral of the square of the fully symmetric amplitudes over the
Dalitz plot (taking care to avoid sextuple counting). Doing this gives
\bea
\mathcal{B}(B^+ \to K^+\pi^+\pi^-)_{fully~sym} = 0.19~\mathcal{B}(B^+ \to K^+\pi^+\pi^-)~, \nn\\
\mathcal{B}(\bd \to K^+ K^0 K^-)_{fully~sym} = 0.50~\mathcal{B}(\bd \to K^+ K^0 K^-)~. 
\eea
{}From Ref.~\cite{hfag}, we have $\tau_0/\tau_+ = 0.93$,
$\mathcal{B}(B^+ \to K^+\pi^+\pi^-) = (51.0 \pm 2.9) \times 10^{-6}$
and $\mathcal{B}(\bd \to K^+ K^0 K^-) = (24.7 \pm 2.3) \times
10^{-6}$.  Eq.~(\ref{relBR}) then gives
\beq
f_{SU(3)} = 0.17 \pm 0.06 ~.
\label{fSU3value}
\eeq
(This error does not include the errors in the parameters obtained
from the Dalitz-plot analyses of the two decays.) 

We can now put all the pieces together to describe how the fit is to
be performed. The fully symmetric amplitudes for the decays $\bd \to
K^+\pi^0\pi^-$, $\bd \to K^0\pi^+\pi^-$, $\bd \to K^+ K^0 K^-$ and
$\bd \to K^0 K^0 \kbar$ are given in Eq.~(\ref{SU3breakamps}). They
are a function of 10 unknown parameters, including $\gamma$.  The
value of $f_{SU(3)}$ is taken from Eq.~(\ref{fSU3value}). The 11
observables and their errors are computed as described above -- the
(fully symmetric) branching ratios and direct CP asymmetries are given
in Eq.~(\ref{BRAdir}), and the indirect CP asymmetries in
Eq.~(\ref{Aindir}). Note that these are for specific values of
$s_{++}$ and $s_+$. One has a different set of observables for each
$(s_{++},s_+)$ pair. With 10 unknowns and 11 constraints, one can now
perform the fit. This will determine the magnitudes and relative
strong phases of the five effective diagrams, as well as $\gamma$, all
for the chosen values of $(s_{++},s_+)$. This is to be repeated for
each independent $(s_{++},s_+)$ pair\footnote{The two pairs
  $(s_{++},s_+)_1$ and $(s_{++},s_+)_2$ are considered as independent
  if $|{\cal M}_{fully~sym}((s_{++},s_+)_1)|$ and $|{\cal
    M}_{fully~sym}((s_{++},s_+)_2)|$ do not overlap when one takes
  into account the errors on the Dalitz-plot parameters of
  Eq.~(\ref{Kpipiamp}).}.  This has two effects. First, one will be
able to fix the momentum dependence of the diagrams. Second, and more
importantly, since $\gamma$ is momentum independent, one can average
over all the $(s_{++},s_+)$ fits. This will reduce its error, perhaps
considerably.

Now, we already have experimental information about most of the
required $\btokpipi$ and $\btokkk$ decays. In particular, the
measurements of the Dalitz plots of $\bd \to K^+\pi^0\pi^-$, $\bd \to
K^0\pi^+\pi^-$ and $\bd \to K^+ K^0 K^-$ are described in
Refs.~\cite{K+pi0pi-}, \cite{K0pi-pi+}, and \cite{K+K-K0},
respectively. On the other hand, we do not yet have the Dalitz plot of
$\bd \to K^0 K^0 \kbar$. The branching ratio and CP asymmetries of
$\bd\to\ks\ks\ks$ are given in Ref.~\cite{KSKSKS}. While the use
of the final state $\ks\ks\ks$ is excellent -- it is proportional to
the fully symmetric state of $K^0 K^0 \kbar$ -- the observables are
momentum independent. That is, an integration over the Dalitz plot has
been performed. However, the method described in this paper requires
the momentum-dependent observables. Once the Dalitz plot for
$\bd\to\ks\ks\ks$ is known, this method for extracting $\gamma$ can be
carried out.

Even though all the experimental data is not yet available, we can
still attempt to estimate the precision with which $\gamma$ can be
obtained. Consider first $\bd \to K^+ K^0 K^-$. According to the BaBar
measurement in Ref.~\cite{K+K-K0}, the largest contributions to this
decay come from the $\phi K^0$ and $f_0 K^0$ resonances, and the $(K^+
K^-)_{NR} K^0$, $(K^+ K^0)_{NR} K^-$ and $(K^- K^0)_{NR} K^+$
non-resonant pieces. They find
\bea
\phi K^0 & : & c_j = 0.0085 \pm 0.0010 ~, \nn\\
f_0 K^0 & : & c_j = 0.622 \pm 0.046 ~, \nn\\
(K^+ K^-)_{NR} K^0 & : & c_j = 1~({\rm fixed}) ~, \nn\\
(K^+ K^0)_{NR} K^- & : & c_j = 0.33 \pm 0.07 ~, \nn\\
(K^- K^0)_{NR} K^+ & : & c_j = 0.31 \pm 0.08  ~,
\eea
where $c_j$ is defined in Eq.~(\ref{Kpipiamp}). The errors, which are
statistical only, range from 7\% to 25\%.  The above method describes
how to obtain ${{\cal M}_{fully~sym}(\bd \to K^+ K^0 K^-)}$ from the
amplitude given in Ref.~\cite{K+K-K0}, and from this the $\bd \to K^+
K^0 K^-$ observables.  A full numerical analysis is needed to do this,
properly taking into account the errors on the $c_j$ above, as well as
the errors on the $\theta_j$ and $F_j$ of Eq.~(\ref{Kpipiamp}), and
the other resonances.  However, a rough guess is that the errors on
the observables will be about $20\%$. Similarly, we (guess)timate that
the errors on the observables of the other decays, including those of
$\bd \to K^0 K^0 \kbar$, will be $\sim 20\%$. In order to obtain
$\gamma$, a fit to the observables must be performed, taking into
account the SU(3)-breaking factor of Eq.~(\ref{fSU3value}) (the error
on $f_{SU(3)}$ will increase once the errors in the Dalitz-plot
parameters are included), and one must average over the independent
$(s_{++},s_+)$ pairs. It is impossible to predict with any accuracy
what the error on $\gamma$ will be, but an error of $O(25\%)$ does not
seem unreasonable. 

How does this compare with the precision on $\gamma$ measured in
two-body decays? The answer is: not that badly. The standard way of
directly measuring $\gamma$ uses $B \to D^0/{\bar D}^0 K$ decays
within the GLW \cite{GLW} or ADS \cite{ADS} methods. The latest
measurement yields $\gamma = (68^{+10}_{-11})^\circ$ \cite{CKMfitter},
i.e.\ the error is $\sim 15\%$. To be sure, our estimated error of
$O(25\%)$ on the value of $\gamma$ as extracted from three-body decays
is worse than $15\%$.  However, it is still roughly the same size, and
if a full analysis were done, the real error might turn out to be
smaller than our estimate. More to the point, when the Dalitz-plot
measurements are done at the super-$B$ factory, the Dalitz-plot
parameters will be obtained with a smaller statistical error. This
will have two effects.  First, the error on $\gamma$ will be reduced
for each $(s_{++},s_+)$ pair. Second, one will have more independent
$(s_{++},s_+)$ pairs, so the error will be further reduced when one
averages over all the $(s_{++},s_+)$ fits. (See the discussion
following Eq.~(\ref{fSU3value}).) Thus, the extraction of $\gamma$
from three-body $B$ decays may turn out to be more precise than that
from two-body decays.

\bigskip
\noindent
Note added: after this paper was submitted, the Dalitz-plot analysis
of $\bd\to\ks\ks\ks$ was submitted to the arXiv, see
Ref.~\cite{KSKSKS}.

\bigskip
\bigskip
\noindent
{\bf Acknowledgments}: Compared to its original version, this paper
has been considerably modified with the addition of the detailed
discussion of how the fit is done, and the guesstimate of the error on
the extracted value of $\gamma$. We are grateful to Jim Smith for
asking the key question which led to this revision.  This work was
financially supported by NSERC of Canada.



\end{document}